\documentclass[aps,prd,showpacs,eqsecnum,nofootinbib]{revtex4}

\usepackage{epsfig}
\usepackage{graphicx}
\usepackage{dcolumn}
\usepackage{amsmath}
\usepackage{enumerate}

\begin{document}

\title{%
\vskip-6pt \hfill {\rm\normalsize UCLA/05/TEP/10} \\
\vskip-12pt~\\
Compatibility of DAMA Dark Matter Detection with Other Searches}

\author{
\mbox{Paolo Gondolo$^{1}$}
and
\mbox{Graciela Gelmini$^{2}$}}

\affiliation{
\mbox{$^1$ Department of Physics, University of Utah,
   115 S 1400 E \# 201, Salt Lake City, UT 84112, USA }
\mbox{$^2$  Department of Physics and Astronomy, UCLA,
 405 Hilgard Ave. Los Angeles, CA 90095, USA}
\\
{\tt paolo@physics.utah.edu},
{\tt gelmini@physics.ucla.edu}}

\date{April 1, 2005}

\vspace{6mm}
\renewcommand{\thefootnote}{\arabic{footnote}}
\setcounter{footnote}{0}
\setcounter{section}{1}
\setcounter{equation}{0}
\renewcommand{\theequation}{\arabic{equation}}

\begin{abstract} \noindent
  We present two examples of velocity distributions for light dark
  matter particles that reconcile the annual modulation signal
  observed by DAMA with all other negative results from dark matter
  searches. They are: (1) a conventional Maxwellian distribution for
  particle masses around 5 to 9~GeV; (2) a dark
  matter stream coming from the general direction of Galactic rotation
  (not the Sagittarius stream).  Our idea is based on attributing the
  DAMA signal to scattering
  off Na, instead of I, and can be tested in the immediate future by
  detectors using light nuclei, such as CDMS-II (using Si) and
  CRESST-II (using O).
\end{abstract}


\maketitle

The nature of dark matter is one of the fundamental problems of
physics and cosmology.  Popular candidates for dark matter are weakly
interacting massive particles (WIMPs).  Direct searches for dark
matter WIMPs aim at detecting the scattering of WIMPs off of nuclei in
a low-background detector. These experiments measure the energy of the
recoiling nucleus, and are sensitive to a signal above a
detector-dependent energy threshold \cite{review}.

One such experiment, the DAMA collaboration \cite{DAMA}, has found an
annual modulation in its data compatible with the signal expected from
dark matter particles bound to our galactic halo \cite{FreeseDrukier}.
Other such experiments, such as CDMS \cite{CDMS-I,CDMS-II}, EDELWEISS
\cite{EDELWEISS,EDELWEISSfinal}, 
and CRESST \cite{CRESST-I,CRESST-II}, have not found
any signal from WIMPs. It has been difficult to reconcile a WIMP
signal in DAMA with the other negative results~\cite{previous}.

Here we show that it is possible to have a dark matter signal above
the WIMP speed threshold for DAMA and below the WIMP speed threshold
for CDMS and EDELWEISS, so that the positive and negative detection
results can be compatible.  We find: (1) that with the standard dark
halo model there is a solution for WIMP masses about 6-9 GeV and
WIMP-proton scattering cross section of about 1 femtobarn
($10^{-39}$~cm$^2$), and (2) that this region of solutions can be
enlarged if a dark matter stream is suitably added to the standard
dark halo. The region in point (1) could certainly also be enlarged by
considering more general halo models, even in the absence of dark
matter streams (see e.g.\ the models in~\cite{DAMA03}).

Light neutralinos as WIMPs with masses as low as 2~GeV
\cite{stodolsky} or, with updated bounds, 6~GeV \cite{bottino} have
been considered, but their cross sections are about one order of
magnitude smaller than those needed here.  In this paper we proceed in
a purely phenomenological way in choosing the WIMP mass and cross
section, although we concentrate on spin-independent cross sections
only. We do not attempt to provide an elementary particle model to
support the values of masses and cross sections.  As justification of
our approach, let us recall that there is no proven particle theory of
dark matter.  The candidates we are considering are stable neutral
particles which have very small cross sections with nucleons, of the
order of femtobarns.  Regarding their production in accelerators, they
would escape from the detectors without interacting.  Unless there is
a concrete specific model relating our neutral candidate to other
charged particles (which yes can be observed) there is no way such
particles could be found in accelerators. The usual signature searched
for in accelerators, for example at LEP, Tevatron or LHC, is the
emission of a charged particle related to the neutral particle in
question.  For example, searching for ``neutralinos" one puts bounds
on one of its cousins, a ``chargino", or another relative, a
``slepton". Without a detailed model there are no accelerator bounds
on neutral dark matter candidates.

\section{Basic idea}

Our idea is that
WIMPs with velocities smaller than the CDMS threshold but larger that
the DAMA threshold could explain the data. Our idea is based on the
following observation.

The minimum WIMP speed required to produce a nuclear
recoil energy $E$ is given by elementary kinematics as
\begin{equation}
v=\sqrt{\frac{ME}{2\mu^2}} = \sqrt{\frac{(m+M)^2 E}{2 M m^2}~.}
\label{v1}
\end{equation}
Here $\mu = m M/(m+M)$ is the reduced WIMP-nucleus mass, $m$ is the
WIMP mass and $M$ is the nucleus mass.  The nuclear energy threshold
$E_{\rm thr}$ observable with a particular nucleus corresponds through
Eq.~(\ref{v1}) to a minimum observable WIMP speed, the speed threshold
$v_{\rm thr}$.  Speed thresholds for several direct detection
experiments listed in Table~1 are plotted in Fig.~1 as a function of
the WIMP mass in the range $m<10$ GeV.  Using Eq.~(\ref{v1}), it is
easy to see that the speed threshold of Na in DAMA is smaller than
that of Ge in CDMS-SUF  for $m < 22.3~{\rm GeV}$.

To understand the dependence of the speed threshold on nuclear mass,
consider the simple case $m \ll M$. Then $\mu \simeq m$ is independent
of the nucleus mass $M$, and $v_{\rm thr}$ is proportional to $\sqrt{
  ME_{\rm thr}}$.  Using the nuclear masses of Na and Ge, $M_{\rm Na}=
21.41$~GeV and $M_{\rm Ge}= 67.64$~GeV, and the energy thresholds in
Table~1, the product $M E_{\rm thr}$ is smaller for Na in DAMA than
for Ge in CDMS-SUF (notice that DAMA used ``electron-equivalent''
energies, which we indicate with keVee units; these need to be
converted into nucleus recoil energies using the so-called quenching
factors listed in the caption of Table~1).  For $m\ll M$, the Ge
$v_{\rm thr}$ in CDMS-SUF is 2.44 times the Na $v_{\rm thr}$ in DAMA.

For $m$ not necessarily much smaller than $M$, we can refer to Fig.~1.
The speed threshold of Ge in CDMS-SUF, of Si and Ge in CDMS-II, of Ge in
EDELWEISS, as well as those of other experiments using heavier nuclei,
are larger than the speed threshold of Na in DAMA in the WIMP mass
range shown.  Three light nuclei, namely Si in CDMS and Al and O in
CRESST, have speed thresholds lower than Na in DAMA, and can be used
to test and constrain our idea.

\begin{figure}[t]
\includegraphics[width=0.45\textwidth]{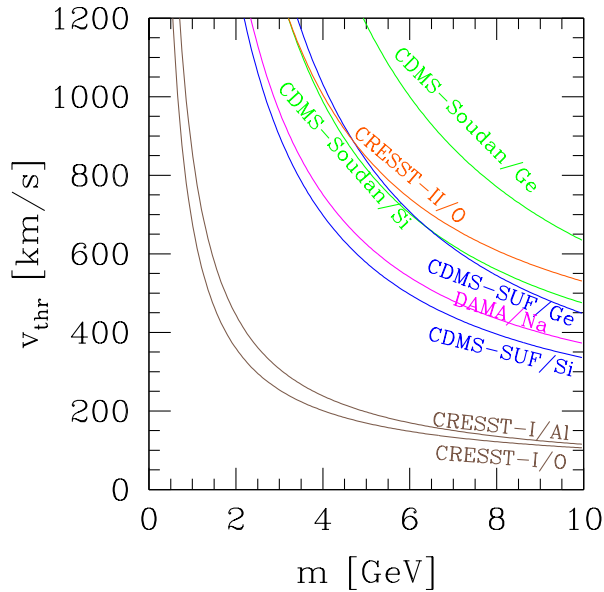}
\caption{ Threshold speeds $v_{\rm thr}$ of several experiments and target 
nuclei. The DAMA Na threshold is lower than the CDMS-SUF Ge threshold for
 $m < 22.3$ GeV.}
\end{figure}

A small component of Si is present in CDMS.  Si is lighter than Ge, although
heavier than Na, $M_{\rm Si}= 26.16$~GeV. Given the nuclear energy
recoil thresholds in Table~1, the speed threshold of Si in CDMS-SUF is
smaller than that of Na in DAMA for all WIMP mass values.  However,
considering the CDMS-SUF efficiency close to 5~keV energies is about
8\%, the effective exposure of the CDMS-SUF Si detector near threshold
is about 0.5~kg-day, which may be too small to have detected the
signal which DAMA might have seen in its Na detector.  In any event,
CDMS has not yet used its Si component to set limits on dark matter,
but only to help in background rejection.

Light nuclei are used by CRESST, in particular O ($M_{\rm O}=
14.90$~GeV).  CRESST-I~\cite{CRESST-I} used sapphire (Al$_2$O$_3$),
which besides O contains Al, similar in mass to Si. CRESST-I has set
limits on dark matter with a very low nuclear recoil threshold of
0.6~keV, but with a small exposure of only 1.5~kg-day. The speed
threshold for O in CRESST-I is so low that CRESST-I is sensitive to
the bulk of the halo dark matter particles we are proposing.
CRESST-II uses calcium tungstate (CaWO$_4$), which also contains the
light O nucleus, but background discrimination sets a relatively high
threshold of $\sim 10$ keV. CRESST-II has run a prototype without
neutron shield and set the limits quoted in Table~1 \cite{CRESST-II}.
The completed CRESST-II will test our idea.

In summary, for light enough WIMPs it could be possible to have dark
matter WIMPs with a speed above threshold for Na in DAMA, and below
threshold for Ge in CDMS and EDELWEISS. That is, we could have a dark
matter signal visible for DAMA but not observable in CDMS and
EDELWEISS and compatible with all experimental data.

\section{Method}

Our procedure is the following. Given a dark halo model, including the
WIMP velocity distribution, we find the viable region by making sure
that we produce the correct amplitude for the DAMA modulation in the
viable region and that all of the current experimental constraints are
satisfied.

We consider constraints from DAMA/NaI-96~\cite{DAMA96},
DAMA/NaI-03~\cite{DAMA03}, EDELWEISS~\cite{EDELWEISS}, 
CDMS-SUF~\cite{CDMS-I}, CDMS-Soudan \cite{CDMS-II},
CRESST-I~\cite{CRESST-I}, and CRESST-II~\cite{CRESST-II}.  The
experimental exposures, efficiencies, thresholds, and constraints we
use are listed in Table~1.

\begin{table*}
\begin{tabular}{||l|l|l|l|l|c||}
\hline\hline
Experiment & Exposure [kg-day] & Threshold [keV] & Efficiency [\%] & Constraint & Ref. \\
\hline
CDMS-SUF & \begin{minipage}{50pt} \begin{flushleft} Si: 6.58\\Ge: 65.8\end{flushleft} \end{minipage} & 5 &  \begin{minipage}{70pt} \begin{flushleft} $E<10$keV: 7.6 \\ $E<20$keV: 22.8 \\ $E>20$keV: 38 \end{flushleft} \end{minipage} & 5--55keV: $<$2.3 events ($\dagger$) & \protect\cite{CDMS-I} \\
\hline
CDMS-Soudan & \begin{minipage}{50pt} \begin{flushleft}Si: 5.26 ($\oplus$)\\Ge: 52.6 \end{flushleft} \end{minipage} & 10 &  \begin{minipage}{70pt} \begin{flushleft}  $E<20$keV: 2.3$E$/keV-8 \\ $E>20$keV: 43.75+$E$/16keV \end{flushleft} \end{minipage} & 10--100keV: $<$2.3 events ($\dagger$) & \protect\cite{CDMS-II} \\
\hline
EDELWEISS& Ge: 8.2 ($\otimes$) & 20 &  100 & 20--100keV: $<$2.3 events  ($\dagger$) & \protect\cite{EDELWEISS} \\
\hline
CRESST-I & Al$_2$O$_3$: 1.51 & 0.6 &  100 &  ($\ddag$) & \protect\cite{CRESST-I} \\
\hline
CRESST-II & CaWO$_4$: 10.448 & 10 &  100 & \begin{minipage}{180pt} \begin{flushleft} Ca+O, 15--40keV: $<$6 events  \\ W, 12--40keV: $<$2.3 events ($\dagger$) \end{flushleft} \end{minipage}  & \protect\cite{CRESST-II} \\
\hline
DAMA/NaI-96 & NaI: 4123.2 & \begin{minipage}{50pt} \begin{flushleft} I: 22 ($\diamond$) \\ Na: 6.7 ($\diamond$)  \end{flushleft} \end{minipage}  &  100 &  \begin{minipage}{180pt} \begin{flushleft} 1--2keVee: $<$1.4/kg-day-keVee ($\star$) \\ 2--3keVee: $<$0.4/kg-day-keVee ($\star$)  \end{flushleft} \end{minipage}  & \protect\cite{DAMA96} \\
\hline
DAMA/NaI-03 & NaI: 107731 & \begin{minipage}{50pt} \begin{flushleft} I: 22 ($\diamond$) \\ Na: 6.7 ($\diamond$)  \end{flushleft} \end{minipage} &  100 & \begin{minipage}{180pt} \begin{flushleft} 2--4keVee: 0.0233$\pm$0.0047/kg-day-keVee ($\bullet$) \\ 2--5keVee: 0.0210$\pm$0.0038/kg-day-keVee ($\bullet$) \\ 2--6keVee: 0.0192$\pm$0.0031/kg-day-keVee ($\bullet$)\\ 6--14keVee: -0.0009$\pm$0.0019/kg-day-keVee($\bullet$) \end{flushleft} \end{minipage} & \protect\cite{DAMA03} \\
\hline\hline
\end{tabular} 
\caption{Experimental constraints used in this study. Notes to the table:
 ($\dagger$) upper limit assuming no detected event; 
 ($\oplus$) only one Si detector is used, the other having ${}^{14}$C contamination \protect\cite{CDMS-I,CDMS-II,Richard}; the Si efficiency is assumed to be the same as the Ge efficiency, for which we take a simple analytic approximation to the curve in Fig.~3 of \protect\cite{CDMS-II};
($\otimes$) final EDELWEISS-I results~\protect\cite{EDELWEISSfinal} with 62 kg-days exposure do not give more stringent bounds because 59 events, attributed to background, have been detected between 10 and 200 keV;
($\ddag$) to reproduce the published curve in~\protect\cite{CRESST-I}, we
 impose appropriate upper limits all along the recoil spectrum in their Fig.~1;
($\diamond$) from an electron
equivalent threshold of 2~keVee,  using the quenching factors
 $Q=E_{\rm ee}/E$ equal to 0.09 for I and  0.3 for Na~\protect\cite{DAMA03};
($\star$) approximations that reproduce the published 
$\sigma_{\rm p}$ vs.\ $m$ limit across our mass range; 
($\bullet$) amplitude of annual modulation from the model-independent 
fit in~\cite{DAMA03} assuming a period of 1 yr and the maximum counting
 rate at June 2.}
\vspace{-10pt}
\end{table*}

To compute the number of recoil events in a given detector we start by
defining the effective exposure of each nuclear species $i$ in the
detector (expressed in kg-days of isotope $i$) as
\begin{equation}
{\cal E}_i = {\cal M}_i T_i \epsilon_i(E) ,
\end{equation}
where $T_i$ is the active time of the detector during which a mass
${\cal M}_i$ of nuclei of species $i$ is exposed to the signal, and
$\epsilon_i(E)$ is the counting efficiency for nuclear recoils of
energy $E$ (for the counting efficiency we assume the values in
Table~1). Then the expected number of recoil events with recoil energy
in the range $(E_1,E_2)$ is the following sum over the nuclear species
in the detector,
\begin{equation}
\label{eq:N}
N_{\text{$E_1$-$E_2$}} = 
\sum_i \int_{E_1}^{E_2} \frac{dR_i}{dE} \, {\cal E}_i(E) \, d E.
\end{equation}
Here $dR_i/dE$ is the expected recoil rate per unit mass of species
$i$ per unit nucleus recoil energy and per unit time. It is
\begin{equation}
\label{eq:3}
\frac{dR_i}{dE} = 
\frac{ \rho \sigma_{i} | F_i(E) |^2 }{ 2 m \mu_{i}^2 } \int_{v \!>\!
 \sqrt{M_i E/2 \mu_{i}^2 }} \frac{f({\bf v},t)}{v} \, d^3v .
\end{equation}
In Eq.~(\ref{eq:3}), $M_i$ is the mass of a nucleus of species $i$,
$m$ is the WIMP mass, $\mu_i = m M_i/(m+M_i)$ is the reduced
WIMP-nucleus mass, $\rho$ is the local halo WIMP density, $F_i(E)$ is
a nuclear form factor for species $i$ (see below), $\sigma_i$ is the
WIMP-nucleus cross section, ${\bf v}$ is the velocity of the WIMP with
respect to the detector, $v=|{\bf v}|$, and $f({\bf v},t)$ is the WIMP
velocity distribution in the reference frame of the detector.

In this analysis, we assume that the WIMP-nucleus interaction is
spin-independent. We make the usual assumption~\cite{review} that the
cross section $\sigma_i$ scales with the square of the nucleus atomic
number $A_i$. Thus, in terms of the WIMP-proton cross section
$\sigma_{\rm p}$, the scaling is $\sigma_i = \sigma_{\rm p} A_i^2
(\mu_i/\mu_{\rm p})^2$.  For the nuclear form factor we use the
conventional Helmi form~\cite{review}, $F_i(E) = 3 e^{-q^2s^2/2}
[\sin(qr)- qr\cos(qr)]/(qr)^3,$ with $s=1$~fm, $r=\sqrt{R^2-5s^2}$,
$R=1.2 A_i^{1/3}$~fm, $q=\sqrt{2 M_i E}$.
 
A technical point: DAMA obtains the nuclear recoil energy $E$ not
directly but as a multiple of a measured electron-equivalent energy
$E_{\rm ee} = Q E$. The quenching factor $Q$ depends on the nuclear
target and has been found experimentally~\cite{DAMA03} to have the
values $Q_{\rm Na} = 0.3$ and $Q_{\rm I}=0.09$. DAMA results are
quoted in electron-equivalent energy (keVee), and Eq.~(\ref{eq:N})
needs to be modified to
\begin{equation}
N_{\text{$E_{\rm ee,1}$-$E_{ee,2}$}} = 
\sum_i \int_{E_{\rm ee,1}/Q_i}^{E_{\rm ee,2}/Q_i} 
\frac{dR_i}{dE} \, {\cal E}_i(E) \, d E.
\end{equation}

The time dependence of the WIMP distribution function $f({\bf v},t)$
is due to the revolution of the Earth around the
Sun~\cite{FreeseDrukier}.  This gives rise to a modulation in the
expected counting rate with a period of a year. For a conventional
halo, the counting rate varies approximately
sinusoidally~\cite{FreeseDrukier}, while for other halo models, in
particular for models with streams, the time dependence is in general
not sinusoidal~\cite{GelminiGondolo}. We define the amplitude of the
annual modulation as half of the difference between maximum and
minimum counting rates. Explicitly, for the $(E_{\rm ee,1}, E_{\rm
  ee,2})$ electron-equivalent energy interval in DAMA, the modulation
amplitude in counts/kg-day-keVee is
\begin{equation}
\label{eq:mod}
{\cal A}_{\text{$E_{\rm ee,1}$-$E_{\rm ee,2}$}} =
 \frac{1}{2} \left[ {\cal R}^{\rm max}_{\text{$E_{\rm ee,1}$-$E_{\rm
 ee,2}$}}
 - {\cal R}^{\rm min}_{\text{$E_{\rm ee,1}$-$E_{\rm ee,2}$}} \right],
\end{equation}
where $ {\cal R}^{\rm max}_{\text{$E_{\rm ee,1}$-$E_{\rm ee,2}$}}$ and
$ {\cal R}^{\rm min}_{\text{$E_{\rm ee,1}$-$E_{\rm ee,2}$}} $ are the
maximum and minimum values of the counting rate during the course of a
year per kg-day-keVee
\begin{equation}
{\cal R}_{\text{$E_{\rm ee,1}$-$E_{\rm ee,2}$}} =
 \frac{1}{{\cal E}_{\rm DAMA}} \, \frac{N_{\rm ee,2}
 - N_{\rm ee,1}}{E_{\rm ee,2} - E_{\rm ee,1} } .
\end{equation}
Here ${\cal E}_{\rm DAMA} = ({\cal M}_{\rm Na}+{\cal M}_{\rm I}) T$ is
the DAMA exposure as listed in Table~1.

We need to fix the WIMP velocity distribution. In this paper we
consider either a conventional Maxwellian distribution or the
conventional Maxwellian distribution plus a dark matter stream (we
discuss these two choices separately in the next sections).

In both cases, we proceed to vary the WIMP mass and the parameters in
the velocity distribution. For each choice of these variables we find
the range of WIMP-proton cross sections $\sigma_{\rm p}$ that produce
the desired modulation amplitude ${\cal A}$ in DAMA/NaI at the
3$\sigma$ or 90\% confidence level. In detail, let $ {\cal
  A}_{\text{2-6}} $ be the expected modulation in the 2-6 keVee bin of
DAMA/NaI-03, as computed using Eq.~(\ref{eq:mod}), and let $ {\cal
  A}_{\text{2-6}}^{\rm DAMA} \pm \delta {\cal A}_{\text{2-6}}^{\rm
  DAMA} $ be the model-independent experimental fit to the annual
modulation amplitude with a fixed period of 1 yr and maximum rate June
2, as provided by the DAMA collaboration in~\cite{DAMA03} and listed in
Table~1 (i.e.\ $ {\cal A}_{\text{2-6}}^{\rm DAMA} \pm \delta {\cal
  A}_{\text{2-6}}^{\rm DAMA} = 0.0233\pm0.0047$ counts/kg-day-keVee).
Analogously, define the quantities $ {\cal A}_{\text{6-14}} $ and $
{\cal A}_{\text{6-14}}^{\rm DAMA} \pm \delta {\cal
  A}_{\text{6-14}}^{\rm DAMA} $ for the 6-14 keVee energy bin.  To
find the best-fit value of $ \sigma_{\rm p} $ we minimize the quantity
\begin{equation}
\chi^2 = \left( \frac{{\cal A}_{\text{2-6}} - 
{\cal A}_{\text{2-6}}^{\rm DAMA}}
{\delta{\cal A}_{\text{2-6}}^{\rm DAMA}} \right)^2 +
 \left( \frac{{\cal A}_{\text{6-14}} - 
{\cal A}_{\text{6-14}}^{\rm DAMA}}
{\delta{\cal A}_{\text{6-14}}^{\rm DAMA}} \right)^2 .
\end{equation}
This gives us $\chi^2_{\rm min}$ and $\sigma_{\rm p,min}$. We accept
the value $\sigma_{\rm p,min}$ only if
\begin{equation}
\chi^2_{\rm min} < 2.
\end{equation}
Then we determine an $n\sigma$ confidence interval for $\sigma_{\rm
  p}$ by imposing that $ \Delta\chi^2 =\chi^2(\sigma_{\rm p})-
\chi^2_{\rm min} $ satisfies
\begin{equation}
\label{eq:chisq2}
\Delta\chi^2  < n^2 ,
\end{equation}
and a 90\% confidence interval by imposing that
\begin{equation}
\Delta\chi^2 < 2.71 .
\end{equation}
These are the appropriate values of $\Delta\chi^2$ for two data points
(${\cal A}_{\text{2-6}}$ and ${\cal A}_{\text{6-14}}$) and one
parameter ($\sigma_{\rm p}$) (see, e.g., the Statistics Section
in~\cite{RPP}).

We have also replaced the 2-6 keVee bin with the 2-4 keVee bin.  One
obtaines larger regions of $\sigma_p$ satisfying our condition.  Since
the 2-4 and 2-6 keVee bins are correlated, we show results for both.

Our procedure differs from previous theoretical
analyses~\cite{previous} in that we use the modulation amplitudes
provided by DAMA in their model-independent analysis~\cite{DAMA03}
instead of their best-fit values obtained fixing the shape of the
nuclear recoil spectrum to that appropriate for the conventional halo
model~\cite{DAMA96}. The latter best-fit values depend on the
conventional recoil spectrum at the specific best-fit WIMP mass of
DAMA ($52^{+10}_{-8}$ GeV). Thus we must use the model-independent fit
to consider different WIMP masses and non-conventional halo models.

Having found a WIMP-proton cross section that produces the DAMA annual
modulation at $n$-sigma (or 90\%) confidence, we evaluate the expected
number of events in all of the other experiments using
Eq.~(\ref{eq:N}), and compare them with the constraints in Table~1. We
require that less than 2.3 events are predicted for each experiment
that observes no events (this is the 90\% C.L. upper bound). All other
upper bounds in Table~1 are also at 90\% C.L.  We thus determine if
the parameters we choose are compatible with all the experimental
constraints we impose.

We take an aditional step in the case in which a dark matter stream is
added to the conventional halo model. After having followed the
procedure described so far, we determine the minimum and maximum
values of the WIMP mass for which there is a (part of the)
$\sigma_{\rm p}$ confidence interval that produces the DAMA annual
modulation and is allowed by all other experiments at 90\% C.L.

\section{Conventional halo model}

Since the experimental bounds on the candidate mass and cross section
depend on the halo model adopted, all dark matter direct detection
experiments conventionally adopt the same isothermal halo model, to be
able to compare their results. In this section, we adopt the same
conventional halo model, so as not to innovate in this respect. We
make no claim that this is a realistic halo model, but it offers us a
definite benchmark for comparison.

\begin{figure}
\includegraphics[width=0.45\textwidth]{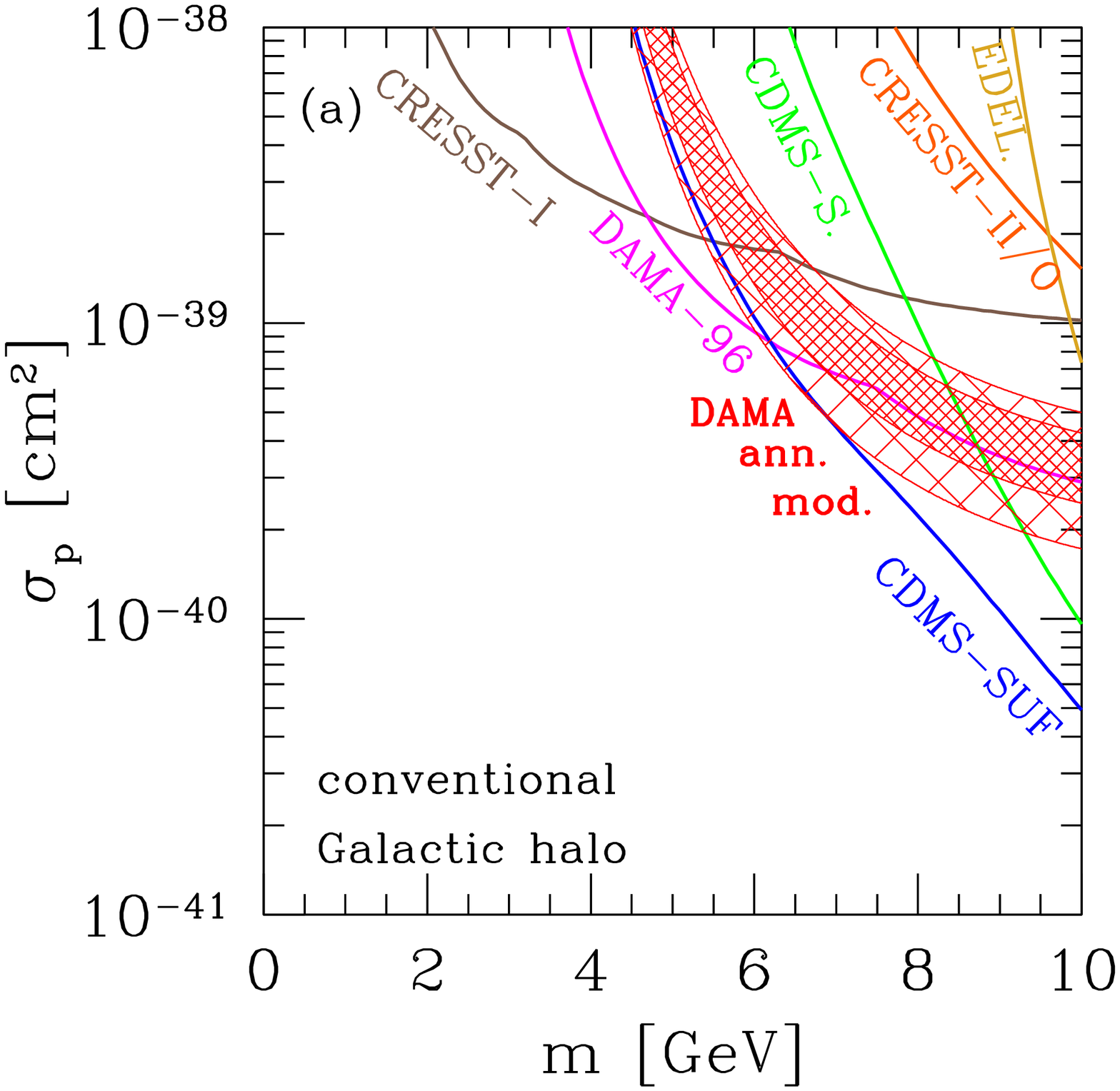}
\includegraphics[width=0.45\textwidth]{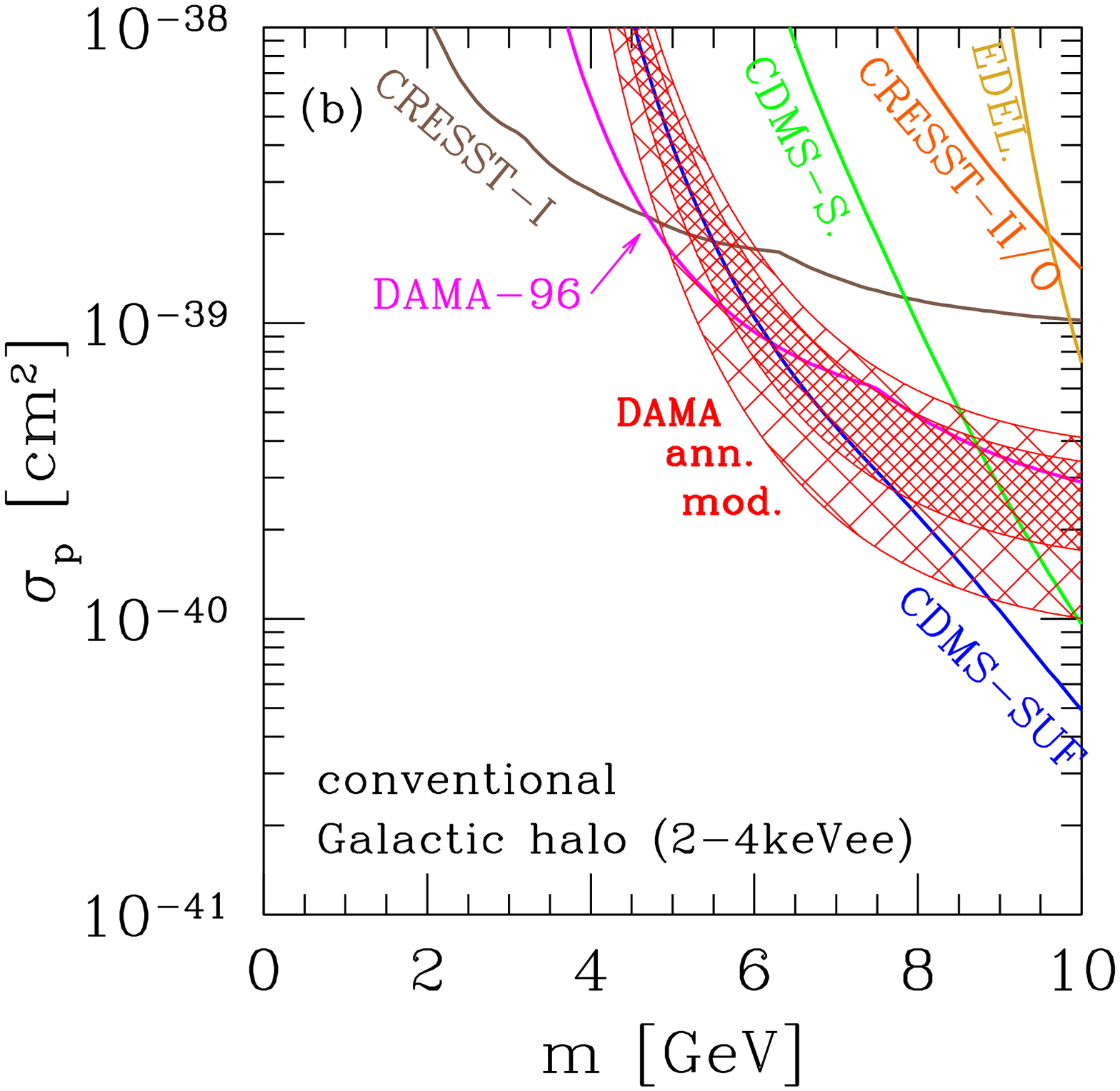}
\caption{Comparison of
  the DAMA annual modulation region with other direct
  detection bounds for spin-independent WIMP-proton interactions and a
  conventional dark halo. In (a) the 2-6 and 6-14 keVee DAMA bins and
  in (b) the 2-4 and 6-14 keVee DAMA bins were used.  In the hatched
  region, the WIMP-proton cross section $\sigma_{\rm p}$ at WIMP mass
  $m$ reproduces the DAMA annual modulation results at the 90\% and
  3$\sigma$ C.L. (inner densely hatched region and outer hatched
  region, respectively).  The region above each other line is excluded
  at 90\% C.L. by the corresponding experiment (DAMA/NaI-96, CRESST-I
  and II, EDELWEISS, 
  CDMS-SUF and CDMS-Soudan [denoted by CDMS-S.]). In (a), there is a region compatible
  with the DAMA annual modulation and all other experiments at the
  3$\sigma$ but not the 90\% C.L. In (b), there is a compatible region
  at the 90\% C.L. also. }
\end{figure}

The value of the local WIMP density conventionally adopted in direct
detection comparisons is $\rho=0.3 $ GeV/cm$^3$. This we adopt.

The conventional WIMP velocity distribution used in the comparison of
direct detection experiments is a Maxwellian distribution truncated at
the local Galactic escape speed $v_{\rm esc}$. In the reference frame
of the detector, which we take to coincide with the reference frame of
the Earth, the conventional WIMP velocity distribution reads
\begin{equation}
\label{eq:hgc}
f_{\rm h}({\bf v},t) = \begin{cases}
 \displaystyle
 \frac{1}{N_{\rm h}(2 \pi {\sigma}_{\rm h}^2)^{3/2}} 
e^{- | {\bf v} + {\bf v}_{\odot} + {\bf v}_{\oplus}(t) |^2/2{\sigma}_{\rm h}^2 } & \text{if 
$|{\bf v} + {\bf v}_{\odot} + {\bf v}_{\oplus}(t) | < v_{\rm esc}$,} \\ 0 & \text{otherwise}.
 \end{cases}
\end{equation}
Here ${\bf v}$ is the velocity of a WIMP relative to the Earth, ${\bf
  v}_{\oplus}(t)$ is the velocity of the Earth relative to the Sun,
and ${\bf v}_{\odot}$ is the velocity of the Sun relative to the
Galactic rest frame, with respect to which the halo WIMPs are assumed
to be stationary. Moreover, $\sigma_{\rm h}$ is the velocity
dispersion of the WIMPs, and $N_{\rm h} = \mathop{\rm
  erf}(z/\sqrt{2})-(2/\pi)^{1/2}ze^{-z^2/2}$, with $z=v_{\rm
  esc}/\sigma_{\rm h}$, is a normalization factor.
 
For ${\bf v}_{\oplus}(t)$, we assume a magnitude of 29.8 km/s and a
direction tangent to a circular orbit on the ecliptic plane.

For ${\bf v}_{\odot}$, we assume the conventional value adopted in our
field, namely 232 km/s in the direction with ecliptic coordinates
$\lambda_{\odot}=340^\circ$, $\beta_{\odot}=60^\circ$.\footnote{This
  velocity of the Sun in the Galactic rest frame is conventional in
  our field. It consists of the IAU recommended value of 220 km/s for
  the galactic rotation speed plus the proper motion of the Sun. In
  reality, estimates of the local galactic rotation speed range from
  170~km/s to 250~km/s (see e.g.~\cite{OllingMerrifield}) and the
  Sun's proper motion is subject to statistical and systematic errors
  of the order of 10 km/s.}

We set the velocity dispersion to the conventional value $
{\sigma}_{\rm h} = 220/\sqrt{2} $ km/s (as applies to an isothermal
model), and we take the escape speed from the galaxy to be $v_{\rm
  esc} = 650$~km/s.  With the halo model we assume, the maximum
possible heliocentric velocity of a halo particle is $v_{\rm esc} +
v_{\odot}= 882$~km/s.

Following the procedure discussed in Section III, we obtain the
results in Figs.~2(a-b). In these figures we show all the experimental
bounds on the WIMP-proton cross section and mass that we obtain using
the content of Table~1, together with the region where the DAMA
modulation is well reproduced at the 90\% C.L. (denser central hatched
region) and 3$\sigma$ C.L. (hatched region).  Figs.~2(a) and~2(b) show
the DAMA region obtained by using their 2-6 and 6-14 keVee data bins
(Fig.~2a) and their 2-4 and 6-14 keVee data bins (Fig.~2b). Also shown
are the regions excluded by the experiments in Table~1 (DAMA/NaI-96,
CRESST-I and II, EDELWEISS, CDMS-SUF, and CDMS-Soudan). 
In Fig.~2(a), there is a region compatible
with the DAMA annual modulation and all other experiments at the
3$\sigma$ but not the 90\% C.L. (notice that the experimental bounds are at the 90\% C.L. and would be less stringent at the 3$\sigma$ level). In Fig.~2(b), there is a compatible region
at the 90\% C.L. also. Since the 2-4 and 2-6 keVee
data bins are correlated, we do not have a means of saying if using
one or the other is better. Thus we conclude that the agreement of
DAMA with all other experiments is marginal when using the standard
halo model. This means, though, that we expect a larger, more
convincing, region of compatibility between DAMA and all the other
experiments if the halo model is extended beyond the conventional
model, for example using the models in Ref.~\cite{DAMA03} or, as
presented in the next section, introducing additional components to
the dark halo, such as a dark matter stream.

\section{Additional dark matter stream}

In this section, we add a stream of dark matter to the conventional
halo model. In this we follow the spirit of~\cite{Sikivie, FGS, GelminiGondolo, stiff}, but look for a stream with
velocity above threshold for Na in DAMA and below threshold for Ge in
CDMS.  Although the stream gives the dominant signal in DAMA (because
there are few halo WIMPs above the DAMA threshold), the signal due to
the bulk of the halo (e.g.\ in CRESST) is not significantly affected.
 
We include the dark matter stream by letting the WIMP velocity
distribution be the sum of the conventional velocity distribution
$f_{\rm h}({\bf v},t)$ of the previous section and a contribution from
the dark matter stream $f_{\rm str}({\bf v},t)$,
\begin{equation}
 f({\bf v},t) = f_{\rm h}({\bf v},t) + f_{\rm str}({\bf v},t).
 \end{equation}
 For the stream contribution, we assume a Gaussian function stationary
 relative to the stream and with velocity dispersion $\sigma_{\rm
   str}$. In the Earth reference frame, the stream velocity
 distribution is
\begin{equation}
f_{\rm str}({\bf v},t) =  \frac{\xi_{\rm str}}{(2 \pi 
\sigma_{\rm str}^2)^{3/2}} e^{- | {\bf v}- {\bf v}_{\rm str} +
 {\bf v}_{\oplus}(t) |^2/2\sigma_{\rm str}^2 } .
\end{equation}
Here $\xi_{\rm str}$ is the local dark matter density in the stream
(in units of the mean local halo density $\rho=0.3$~GeV/cm$^3$), ${\rm
  v}$ is the velocity of the WIMP relative to the Earth, ${\bf
  v}_{\oplus}(t)$ is the velocity of the Earth relative to the Sun,
and ${\bf v}_{\rm str}$ is the velocity of the stream relative to the
Sun (heliocentric velocity).

We fix the stream velocity dispersion $\sigma_{\rm str}=20$~km/s.
Factor-of-2 variations of this value lead to small changes in the results.
We consider stream density fractions $\xi_{\rm str}$ of up to 3\%.
Detailed studies of the evolution of possible residual substructure in
the galactic dark halo concluded that there is a high probability for
the Earth to be passing through a dark matter clump or stream with
density $\sim 3$\% of the mean local halo density~\cite{stiff}.
Ref.~\cite{stiff} studied mostly clumps that have orbited the Galaxy
between one and four times (although mentioning that clumps that have
orbited the Galaxy more times can also produce cold, high velocity
streams in the solar neighborhood) and found plausible clump
velocities relative to Earth of 400 to 700~km/s and velocity
dispersions of 20 to 50~km/s. The dark matter stream may also be of
extragalactic origin. We comment about this possibility later.

Instead of the direction of ${\bf v}_{\rm str}$, we specify the
ecliptic coordinates $(\lambda_{\rm str}, \beta_{\rm str})$ of the
arrival direction of the stream, i.e.\ the direction of $-{\bf v}_{\rm
  str}$. This we do to be able to compare directly with the arrival
direction in the conventional halo model, which is the direction of
motion of the Sun in the Galactic rest frame
$(\lambda_{\odot},\beta_{\odot})=(340^\circ,60^\circ)$.

In order to account for the DAMA annual modulation, the stream arrival
direction is limited by the requirement that the DAMA modulation peaks
May 21 $\pm$ 22 days. To find the direction of the Earth velocity at
that time, we reason as follows. May 21 is 61 days after the Spring
equinox (March 21), and thus the Sun is at ecliptic longitude
$61/365.25 \times 360=60^\circ$. Since the radius vector to the Sun
and the velocity of the Earth are almost perpendicular (they are
exactly so for a circular orbit), May 21 the Earth is moving toward a
point of ecliptic longitude $60^\circ-90^\circ=330^\circ$. If the DAMA
annual modulation is due entirely to a stream, its arrival direction
on Earth should have ecliptic longitude $\lambda=330^\circ \pm
22^\circ$. Since we use the modulation amplitudes obtained by DAMA
assuming a phase such that the maximum rate occurs on June 2, we fix
the ecliptic longitude of the arrival direction of the stream at
$\lambda=340^\circ$.  The amplitude of the modulation depends on the
projection of the stream velocity onto the ecliptic, and is
proportional to $\cos\beta_{\rm str}$, where $\beta_{\rm str}$ is the
ecliptic latitude (i.e. the angle above the plane of the ecliptic) of
the stream arrival direction.

\begin{figure}
\includegraphics[width=0.45\textwidth]{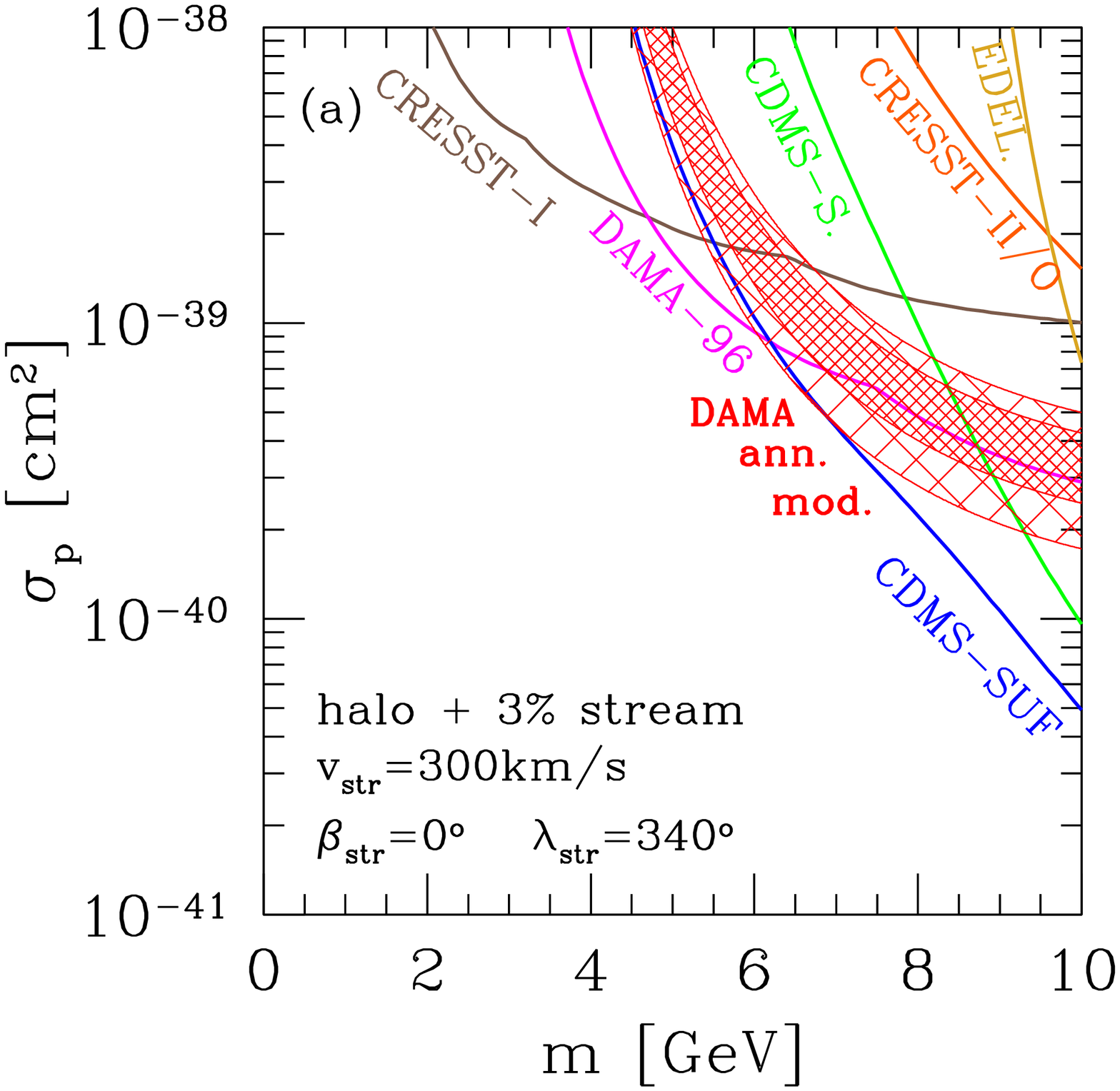}
\includegraphics[width=0.45\textwidth]{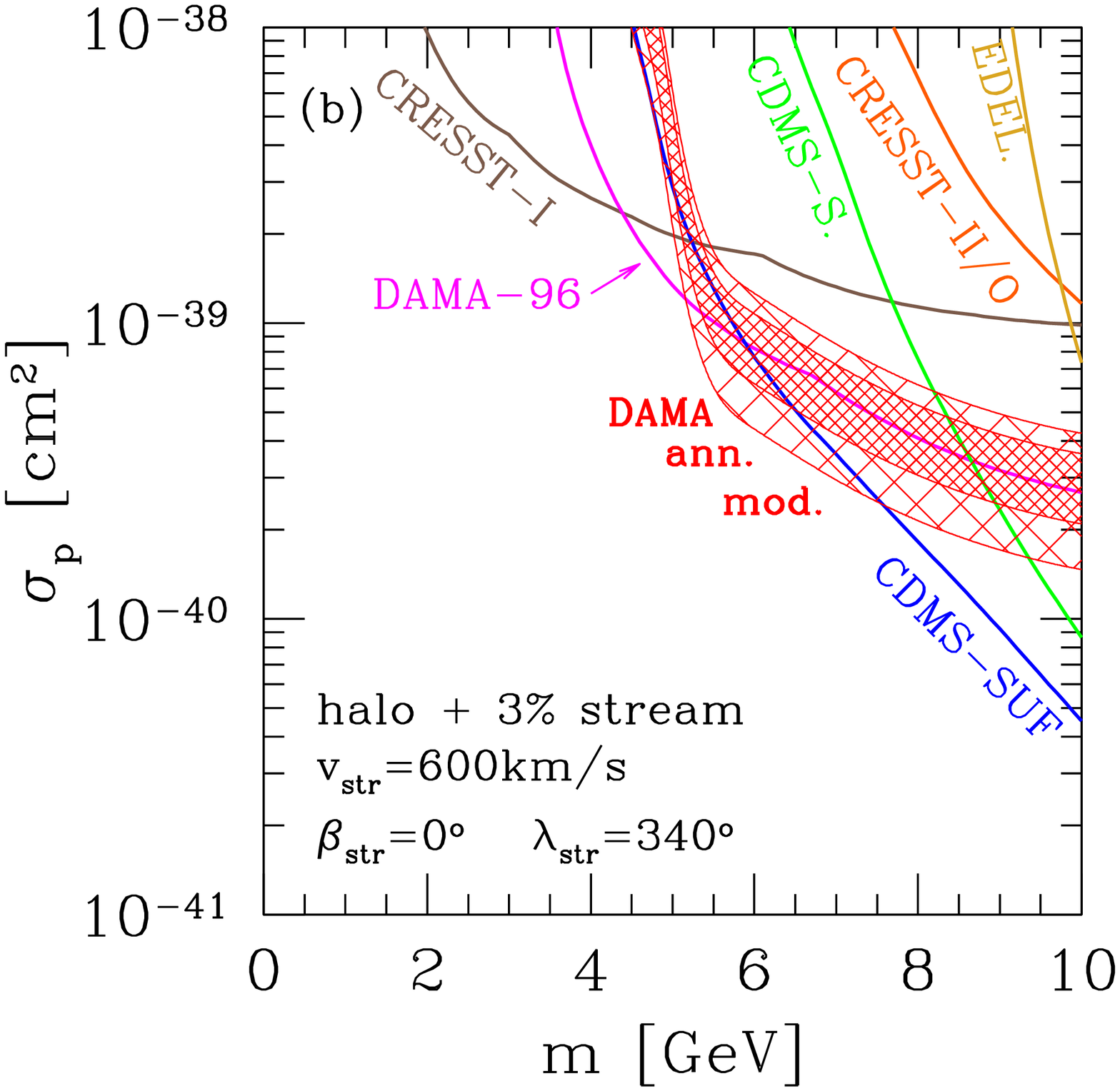}
\includegraphics[width=0.45\textwidth]{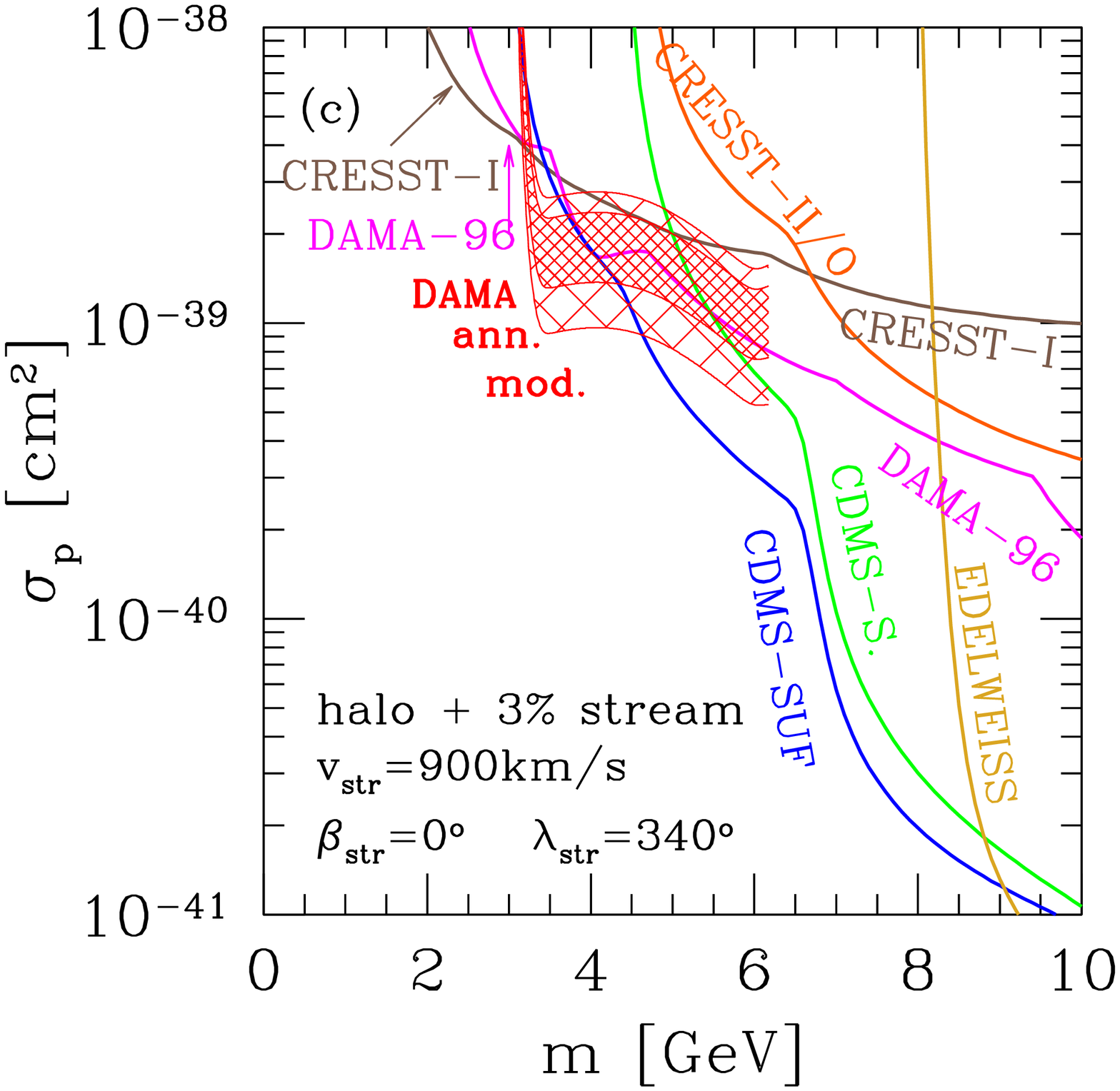}
\includegraphics[width=0.45\textwidth]{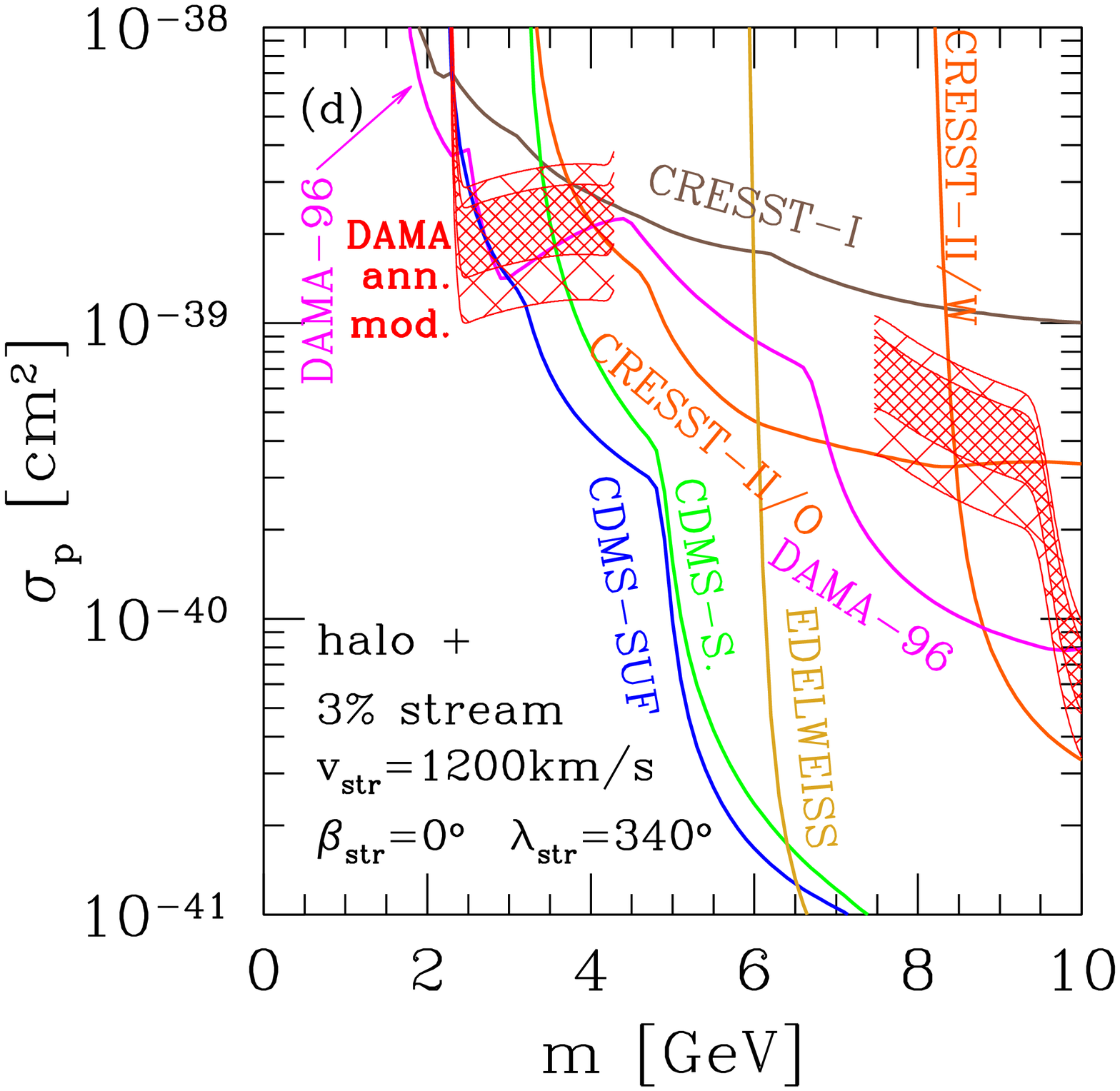}
\caption{Same as Fig.~2(a) but with the addition of a dark matter stream
  with density 3\% of the conventional local halo density,
  heliocentric arrival direction of ecliptic coordinates
  $(\lambda_{\rm str}, \beta_{\rm str})=(340^\circ, 0^\circ)$, and
  heliocentric speed of (a) 300~km/s, (b) 600~km/s, (c) 900~km/s, and
  (d) 1200~km/s. The DAMA modulation region is shown both for the 90\%
  and the 3$\sigma$ C.L. (inner densely hatched and outer hatched
  regions, respectively). The gaps in the DAMA modulation region in
  panels (c) and (d) are due to our requirement that $\chi^2_{\rm
    min}<2$. The experimental upper limits change when the stream is
  included.}
\end{figure}

\begin{figure}
  \includegraphics[width=0.45\textwidth]{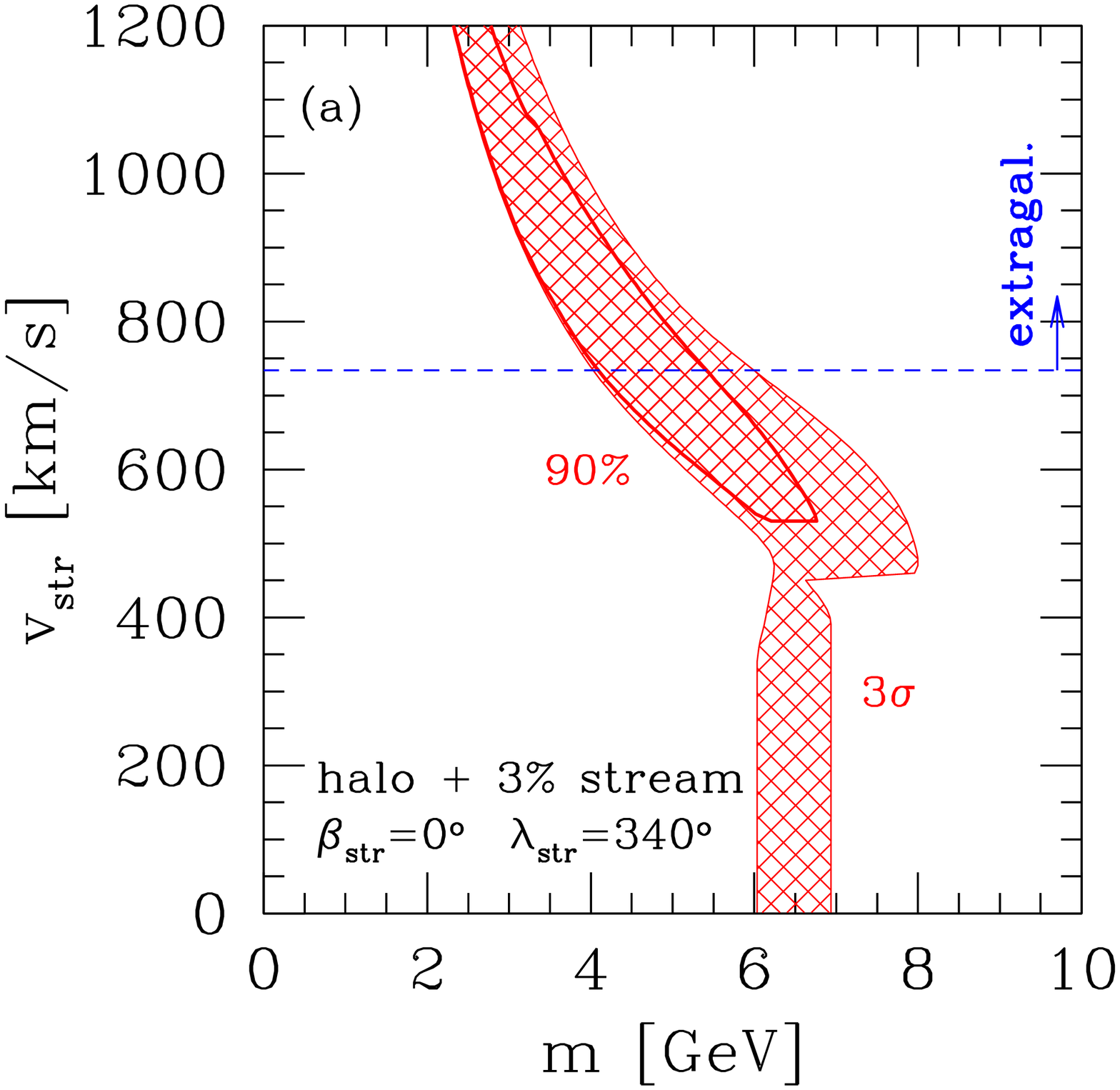}
  \includegraphics[width=0.45\textwidth]{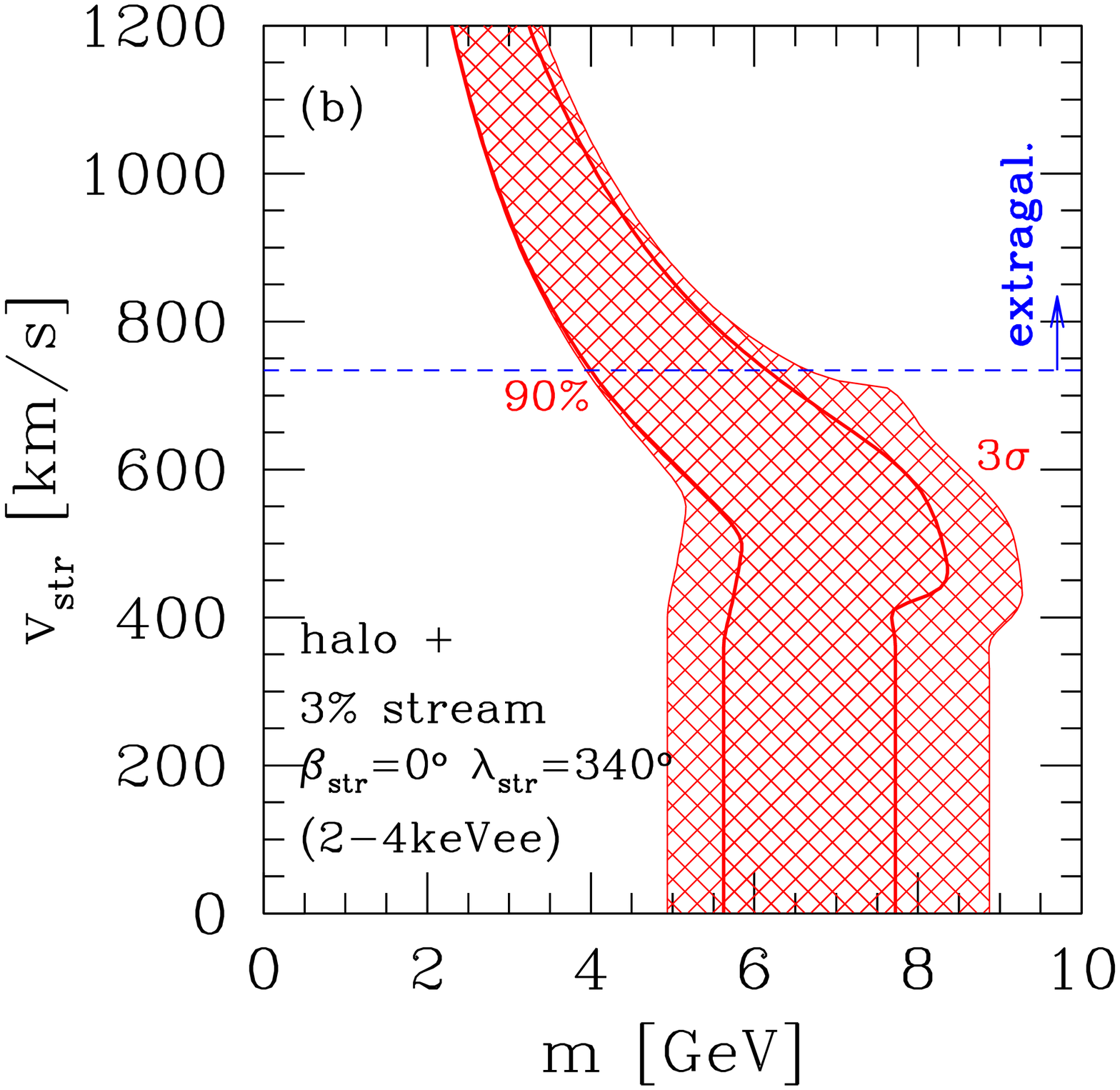}
\caption{Range of WIMP masses $m$ for which there is a 
  compatible region between the DAMA modulation and the other
  experimental results at various stream heliocentric speeds $v_{\rm
    str}$.  Here the stream is assumed to arrive from ecliptic
  longitude $\lambda_{\rm str}=340^\circ$ and ecliptic latitude
  $\beta_{\rm str}=0^\circ$.  Also indicated is the speed above which
  the stream is extragalactic (dashed horizontal line).  The inner
  densely hatched and outer hatched regions correspond to the 90\% and
  3$\sigma$ C.L., respectively.  In (a), we use the 2-6 and 6-14 keVee
  DAMA bins; in (b), the 2-4 and 6-14 keVee bins. At the 3$\sigma$
  level it is possible to find WIMP masses compatible with DAMA and
  all other experiments at any assumed stream speed. }
\end{figure}

\begin{figure}
\includegraphics[width=0.45\textwidth]{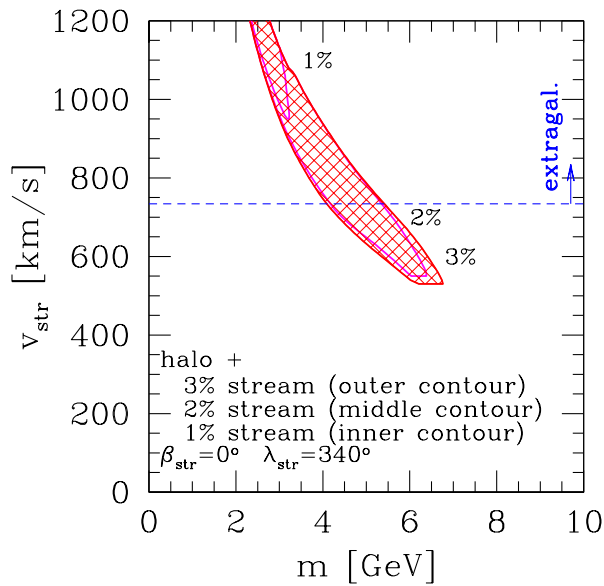}
\caption{Same as Fig.~4(a) at the 90\% C.L. but showing the effect of changing the stream density from 3\% to 2\% and to 1\% of the local smooth halo density.}
\end{figure}

\begin{figure}
  \includegraphics[width=0.45\textwidth]{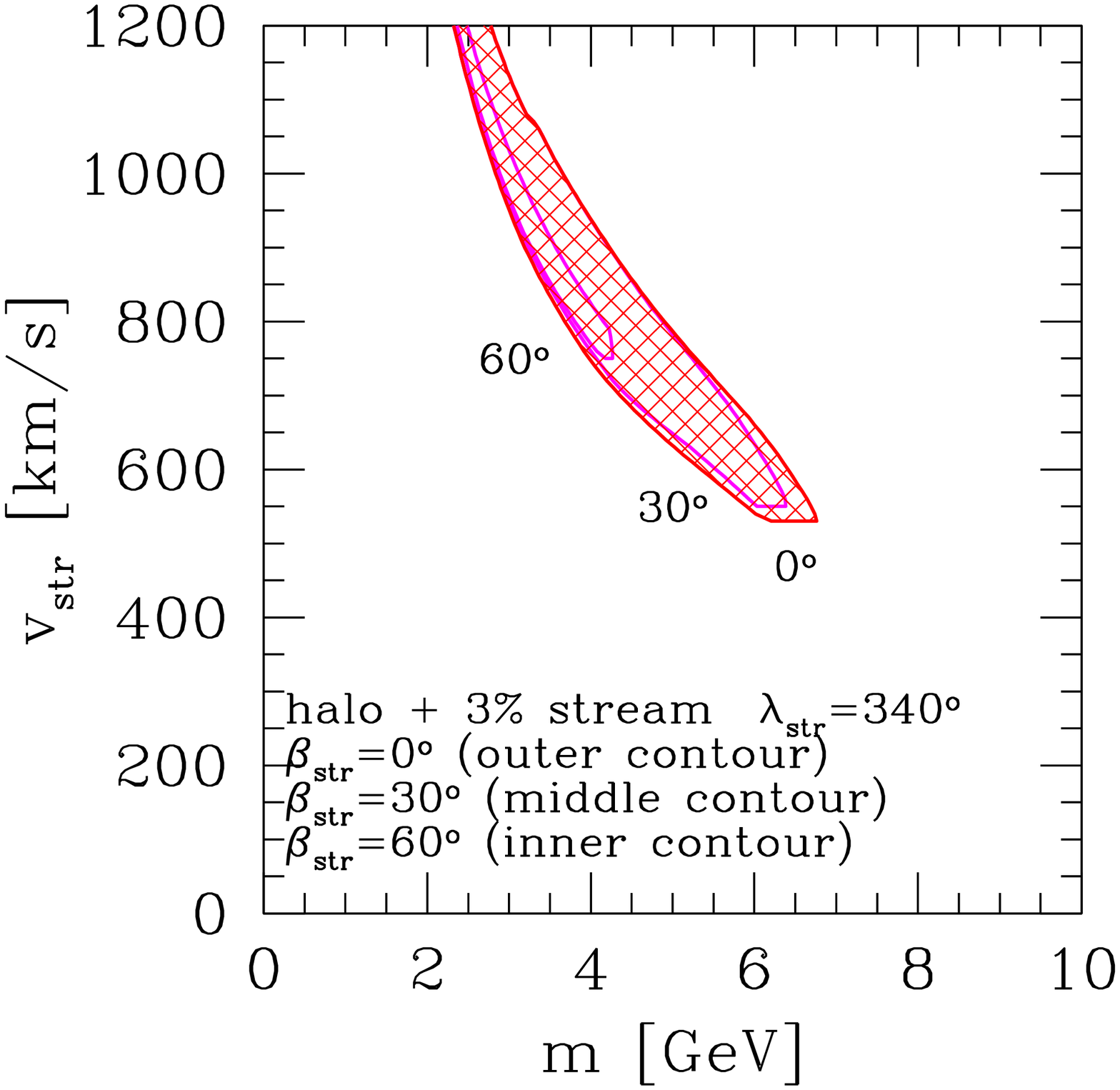}
\caption{Same as Fig.~4(a) at the 90\% C.L. but showing the effect of changing the stream arrival direction from $\beta_{\rm str}=0^\circ$ to $\beta_{\rm str}=30^\circ$ and to $\beta_{\rm str}=60^\circ$.}
\end{figure}

We let the magnitude of ${\bf v}_{\rm str}$ vary between $v_{\rm
  str}=0$ and $v_{\rm str}=1200$~km/s, thus allowing for both Galactic
and extragalactic streams (i.e.\ streams bound or not bound to the
Galaxy). A stream is unbound if its Galactocentric speed $ | {\bf
  v}_{\rm str} + {\bf v}_{\odot} | $ exceeds the local escape speed.
In formulas, the stream is extragalactic if $v_{\rm str} > v_{\odot}
\cos\phi + [ v_{\rm esc}^2 - v_{\odot}^2 \sin^2\phi ]^{1/2} $ where $
\cos\phi = \cos\beta_{\rm str} \cos\beta_{\odot} \cos(\lambda_{\rm
  str}-\lambda_{\odot})$.

Streams bound to our own galaxy have been observed, for example the
tidal streams of the Sagittarius dwarf galaxy \cite{Sgr}. The
Sagittarius leading tidal stream might pass through or close to the
solar neighborhood, with a Galactocentric speed of $\approx 200$~km/s,
but its arrival direction $(\lambda_{\rm Sgr},\beta_{\rm Sgr})\approx
(187^\circ,8^\circ)$ does not have the correct ecliptic longitude to
give the observed phase of the DAMA modulation without a substantial
contribution from the usual halo component. So the stream we are
implying should be a different, perhaps yet undiscovered, stream.

There may be dark matter bound not to our galaxy but to our Local
Group of galaxies~\cite{localg}, and also dark matter bound to our
supercluster, possibly passing through us~\cite{FGS}.  A stream made
of such dark matter has its Galactocentric incoming speed $v_{\rm in}$
at large distances from the Galaxy increased by gravitational focusing
while falling into the Galaxy. Its resulting Galactocentric speed near
the Sun is $v_{\rm local} =\sqrt{v_{\rm in}^2 +v_{\rm esc}^2}$. Its
velocity dispersion may increase by a factor $v_{\rm local}/v_{\rm
  in}$ if the stream self-gravity is negligible, or remain
approximately constant in self-gravitating regions of the stream.  To
give an idea of the Galactocentric velocities involved, a stream with
arrival direction $( \lambda_{\rm str}, \beta_{\rm
  str})=(340^\circ,0^\circ)$ and heliocentric speed $v_{\rm
  str}=800$~km/s requires a local Galactocentric speed $v_{\rm
  local}=712$~km/s, and, accounting for gravitational focusing, a
Galactocentric incoming speed at large distances $v_{\rm in} =
293$~km/s. This value is of the same order of magnitude as the
Galactic speed relative to the Cosmic Microwave Background (549~km/s,
using the CMB dipole measurement in Ref.~\cite{Lineweaver} and the
conventional Galactocentric velocity of the Sun in section IV).

The density of an incoming stream is also increased by focusing, at
least linearly with the ratio $v_{\rm local}/{v_{\rm in}}$ (see the
argument is Ref.~\cite{FGS}), but possibly by much larger factors,
which are however complicated to evaluate. To have a local stream density fraction of the order of a few percent and Galactocentric incoming speeds $v_{\rm in} \sim 100$ km/s, the density of the infalling dark matter at infinity must be of the order of the local halo density. This may be possible if a small dark galaxy bound to our supercluster happens to cross our own.
Dark matter bound to the Local Group of galaxies is expected to have a much smaller density.
The average density of the
Local Group of galaxies is of the order of 2.2 times the critical
density, i.e. 0.6$\times 10^{-5}$ GeV cm$^{-3}$, which is only 2
$\times 10^{-5}$ of the local halo density. In order for a stream with
say 0.1 of the average density far away from our Galaxy to reach a
density of 3\% of the local halo density in the solar neighborhood, a
focusing enhancement factor of 1.5 $\times 10^{4}$ is needed. This
could only be possible if $v_{\rm in}$, the velocity of the stream
with respect to our galaxy, is close to zero, i.e.\ if the stream or
clump of dark matter was initially almost at rest with respect to our
Galaxy (in this case $v_{\rm local} \simeq v_{\rm esc}$). Moreover,
since the velocity dispersion of the stream may be enhanced by a
factor of $10^4$ unless the Solar System happens to be within a
self-gravitating region of the stream, the initial velocity dispersion
may have to be very small too. 

Since the last considerations about an
extragalactic stream are rather speculative, we have decided to
present results with a wide range of stream velocties, up to 1200
km/s.
 
To illustrate the effect of a dark matter stream on the allowed WIMP
mass and cross section, we take a stream density fraction $\xi_{\rm
  str}=0.03$ and a stream arrival direction $(\lambda_{\rm
  str},\beta_{\rm str})=(340^\circ, 0^\circ)$ so that the stream is on
the plane of the ecliptic. In Figs.~3(a-d), we use the 2-6 and 6-14
keVee DAMA bins (the same as in Fig.~2a) to find the region in which
the WIMP-proton cross section $\sigma_{\rm p}$ at WIMP mass $m$
reproduces the DAMA annual modulation results at the 90\% and
3$\sigma$ C.L. (inner densely hatched region and outer hatched region,
respectively). The other lines are experimental upper bounds: the
region above each line is excluded at 90\% C.L. by the corresponding
experiment (DAMA/NaI-96, CRESST-I and II, EDELWEISS, CDMS-SUF, and CDMS-Soudan). Notice
by comparing Figs.~2 and 3 that the experimental upper limits change
when the stream is included.  Figs.~3(a-d) correspond to heliocentric
speeds of (a) 300~km/s, (b) 600~km/s, (c) 900~km/s, and (d) 1200~km/s.
The gaps in the DAMA modulation region in panels (c) and (d) are due
to our requirement that $\chi^2_{\rm min}<2$.  We have arbitrarily
considered stream velocities up to $v_{\rm str} = 1200$~km/s, which is
an extremely high velocity, just to show a complete picture.
Excluding the highest values of $v_{\rm str} $ would eliminate the
compatiblity region at the lowest WIMP masses.
  
The range of WIMP masses $m$ compatible with DAMA and all other
experiments is plotted in Figs.~4(a-b) for varying stream heliocentric
speeds $v_{\rm str}$.  Here we again take a stream density fraction
$\xi_{\rm str}=0.03$ and a stream arrival direction $(\lambda_{\rm
  str},\beta_{\rm str})=(340^\circ, 0^\circ)$.  The dashed horizontal
lines indicate the heliocentric speed above which the stream must be
extragalactic (for the given arrival direction).  The inner densely
hatched and outer hatched regions correspond to the 90\% and 3$\sigma$
C.L., respectively.  In Fig.~4(a), we use the 2-6 and 6-14 keVee DAMA
bins; in Fig.~4(b), the 2-4 and 6-14 keVee bins.  Fig.~4 shows that at
the 3$\sigma$ level it is possible to find WIMP masses in the range of
2 to 9 GeV compatible with DAMA and all other experiments at any
assumed stream speed.

For different values of the stream density and arrival direction, the
results shown in Fig.~4 change.  In Figs.~5 and 6 we show the
variation expected in Fig.~4(a) for the 90\% C.L.  Similar
changes in the size of the compatibility region happen in all other
cases considered.  In Fig.~5 we show the effect of changing the stream
density from 3\% to 2\% and to 1\% of the local smooth halo density.
The compatibility region decreases with decreasing stream density.
Densities of at least 1\% of the local smooth halo density are need to
have a sizable effect.  In Fig.~6 we show the effect of changing the
stream arrival direction from $\beta_{\rm str}=0^\circ$ to $\beta_{\rm
  str}=30^\circ$ and to $\beta_{\rm str}=60^\circ$.  For $\beta_{\rm
  str})= 0^\circ$, the arrival direction of the stream is on the plane
of the orbit of the Earth around the Sun, the plane of the ecliptic.
For $\beta_{\rm str}= 30^\circ$, the stream is at 30$^\circ$ of the
ecliptic and at 30$^\circ$ of the Sun's Galactocentric velocity.  For
$\beta_{\rm str}=60^\circ$, the arrival direction of the steam is
aligned with the Sun's Galactocentric velocity. As $\beta_{\rm str}$
increases the effect of the stream in the modulation decreases,
eventually becoming zero when the stream arrives perpendicularly to
the plane of the ecliptic.

\section{Conclusions}

We have pointed out that for light dark matter particles a signal
could be observed by DAMA through its Na, instead of I, component.
Such a signal would be below threshold for Ge in CDMS and EDELWEISS.
This possibility can be tested with a few months of Si data in
CDMS-Soudan, and future O data in CRESST.

For WIMPs with spin independent interactions, we have presented two
examples of dark matter velocity distributions that give the annual
modulation observed by DAMA but satisfy all other constraints from
dark matter searches.

The first is a conventional Maxwellian distribution with a WIMP mass
around 5 to 9~GeV.  Our results are shown in Fig.~2.  This simple
possibility remains marginally open. This suggests that if several
other possible dark galactic halo models were considered the region of
compatibility would be larger.

Our second example is the conventional distribution superposed to a
dark matter stream coming from the general direction of the Galactic
rotation (but {\it not} the Sagittarius stream).  For the sake of
illustration, we have assumed a particular density of the stream (0.03
of the local halo density) and a particular incoming direction (on the
plane of the Earth's orbit and the direction of the Sun's velocity in
the Galaxy) to obtain the allowed regions presented in Figs.~3 and 4.
Figs.~5 and~6 show how the compatibility regions decrease with
decreasing stream density and as the direction of arrival of the
stream moves away from the plane of the Earth's orbit around the Sun.
The effect of the stream is larger (smaller) for larger (smaller)
stream densities and for incoming directions closer to (further from)
the plane of the Earth's orbit.

For simplicity, we have illustrated our idea only for the case of
WIMPs with spin-independent interactions.  Other kinds of particles
and interactions, or halo velocities distributions more complicated
than a conventional Maxwellian distribution, may extend the allowed
regions of parameters.

\vspace{0.3cm}
{\bf Acknowledgements}

We thank Richard Schnee for extensive valuable discussions on the CDMS detector and results.

This work was supported in part by the US DOE grant DE-FG03-91ER40662 
Task C and NASA grants NAG5-13399  and ATP03-0000-0057.

\end{document}